\documentclass[aps,prl,
twocolumn, showpacs,twoside,amsmath,amsfonts]{revtex4}
\newcommand{\be}{\begin{equation}}
\newcommand{\ee}{\end{equation}}
\newcommand{\bea}{\begin{eqnarray}}
\newcommand{\eea}{\end{eqnarray}}
\newcommand{\e}{\epsilon}
\newcommand{\de}{\partial}
\newcommand{\dw}{\partial_\omega}
\newcommand{\dt}{\partial_\tau}
\newcommand{\hg}{\hat{\Gamma}}
\newcommand{\sech}{\textrm{Sech}}

\newcommand{\espace}{\!\!\!\!\!\!\!\!\!\!}
\newcommand{\w}{\omega}
\newcommand{\li} {\textrm{Li}}
\newcommand{\ha}{{\cal H}}
\usepackage{graphicx}
\usepackage{epsfig}
\usepackage{amssymb}
\usepackage{amsfonts}
\usepackage{amsmath}
\usepackage{graphicx}
\usepackage{psfrag}
\begin{document}
\title{
%Phase coherence and inelastic scattering \\ in pumping through interacting quantum dots.
Phase coherence, inelastic scattering, \\ and interaction
corrections in pumping through quantum dots }
\author{Davide Fioretto$^1$, Alessandro Silva$^2$}
\affiliation{$^1$ Dipartimento di Fisica, Universita' degli Studi di
Milano, via Celoria 16, 20133 Milano, Italy \\ $^2$ The Abdus Salam
International Centre for Theoretical Physics, Strada Costiera 11,
34100 Trieste, Italy}
%\date{Month DD, Year}
\begin{abstract}
Adiabatic quantum pumping in noninteracting, phase coherent quantum
dots is elegantly described by Brouwer's formula. Interactions
within the dot, while suppressing phase coherence, make Brouwer's
formalism inapplicable. In this paper, we discuss the nature of the
physical processes forcing a description of pumping beyond Brouwer's
formula, and develop, using a controlled adiabatic expansion, a
useful formalism to study the effect of interactions within a
generic perturbative scheme. The pumped current consists of a first
contribution, analogous to Brouwer's formula and accounting for the
remanent coherence, and of interaction corrections describing
inelastic scattering. We apply the formalism to study the effect of
interaction with a bosonic bath on a resonant level pump and discuss
the robustness of the quantization of the pumped charge in turnstile
cycles.

%Studying in detail the dependence .. The inclusion of such
%corrections is show to preserve the quantization of the pumped
%charge in turnstile cycles.
%We study adiabatic quantum pumping through interacting quantum dots.
%In particular, we discuss the nature of the physical processes
%forcing a description of pumping beyond Brouwer's formula, showing
%that interaction induced corrections to it are related to inelastic
%scattering events. We demonstrate this by deriving a general formula
%for the corrections within a controlled, systematic gradient
%expansion, and applying the formalism to study the effect of
%interaction with a generic equilibrium bosonic bath on a resonant
%level quantum pump. Finally, we compute the dependence of the pumped
%charge on the bath temperature, and discuss the robustness against
%interaction of charge quantization in turnstile cycles.
\end{abstract}

\maketitle

\it Introduction \rm - Adiabatic quantum pumping in quantum dots
(QD) is one of the simplest, yet non trivial settings where the
interplay between phase coherence and non-equilibrium dynamics can
be studied both theoretically~\cite{Thouless,Brouwer,
Spivak95,Vavilov,AA98,Fazio,Fazio2,Sela} and
experimentally~\cite{exp}. In a quantum pump, a slow, cyclic
variation of the system parameters in time generates a direct
current through the system~\cite{Brouwer}. For suitable cycles the
pumped charge may be quantized~\cite{Thouless,Slava}, a fact that
might have important implications for metrological purposes.

In noninteracting, phase coherent systems, this phenomenon is
encoded elegantly in Brouwer's formula~\cite{Brouwer}, describing
geometrically the charge pumped per cycle in terms of the
instantaneous scattering matrices of the system. Pumping in
interacting systems is much less understood. In the past years
several ad hoc approaches have been developed to address specific
interacting problems~\cite{AA98}. More recently two interesting
studies~\cite{Fazio,Fazio2,Sela} aimed at deriving generalized
pumping formulas, analogous to Brouwer's one, within different
approximation schemes (the average time
approximation~\cite{Fazio,Fazio2}, and linear response~\cite{Sela}).
The formulas so obtained, while in principle equivalent, are both
hardly usable as a starting point to address generic pumping
problems. In addition, the lack of a general understanding of the
physics of interacting quantum pumps makes it difficult to develop
systematic approximation schemes.

The purpose of this paper is twofold. First of all, we develop a
systematic physical understanding of the new elements brought in by
interaction in the physics of quantum pumps. Secondly, by making use
of a controlled gradient expansion~\cite{Vavilov}, we derive a handy
formalism to describe the effect of interactions on pumping within a
generic perturbative scheme. The current  can be conveniently
written as the sum of two terms. The first, analogous to Brouwer's
formula (but expressed in terms of the elastic S-matrices), accounts
the remanent phase coherence. The second term express the
corrections due purely to interaction and contains the new physics
brought in by it: inelastic scattering~\cite{Moskalets}. We
demonstrate this by studying the effect of the coupling to a generic
\it equilibrium \rm environment on a resonant level quantum
pump~\cite{Slava}. Computing the dependence of the pumped charge on
the bath temperature, we prove the robustness against interaction of
the quantization of the pumped charge in turnstile
cycles~\cite{Slava}.

%address two fundamental questions concerning interacting quantum
%pumps: when is a description beyond Brouwer's formula compulsory ?
%And what are the leading physical processes governing the%
%corrections to it ? As shown in the rest of the paper, the crucial
%players in answer to these questions are interaction induced
%inelastic scattering events. We demonstrate this in two steps: first
%using a controlled gradient expansion~\cite{Vavilov} we derive the
%generic form of the corrections to Brouwer's formula. Secondly, we
%apply this formulation to

\it Inelastic scattering and pumping \rm - We begin with a
qualitative discussion of the effects of inelastic scattering on
adiabatic quantum pumping. To illustrate our ideas, we consider a
concrete example of interacting quantum pump: a single level quantum
dot (QD) coupled to a generic bosonic environment in equilibrium at
temperature $T$. The Hamiltonian is \bea &&\espace {\cal
H}(t)=\sum_{k,\alpha} \epsilon_k\, c^\dagger_{k \alpha}\,c_{k
\alpha}+ \epsilon(t)\, d^\dagger\, d+\sum_{q} \omega_q\,
a^\dagger_{q}\,a_{q}+ \nonumber \\
&&\espace + \sum_{k,\alpha}\left[V_\alpha (t)\, d^\dagger\, c_{k \alpha} +
  h.c.\right] +d^\dagger\,d \sum_q \lambda_q \left(a_q^\dagger+a_{-q} \right).
\label{H_ph} \eea Here the $d$'s refer to the resonant level, the
$a_q$'s describe the bosonic modes, while $c_{k \alpha}$ annihilates
an electron in the left/right leads ($\alpha=L,R$). The chemical
potential of the two leads is assumed equal and put to zero. The
instantaneous strength of the coupling to the leads is characterized
by the parameter $\Gamma(t)=\sum_{\alpha}
\Gamma_\alpha=\sum_{\alpha} 2\,\pi\,\nu |V_\alpha (t)|^2$, where
$\nu$ is the density of states in the leads at the Fermi level and
the parameters $V_{\alpha}$ (as well as $\epsilon$) are assumed to
vary cyclically in time with frequency $\Omega$. The environment is
specified through its spectral density \be J(\w)=\sum_q
\lambda_q^2\, \delta (\w-\w_q)=C \, \theta(\w) \,\w^s \,e^{-\frac \w
{\w_c}} , \ee where C is a constant, $\w_c$ a high frequency cutoff,
and $s \lessgtr 1$ corresponds to a sub/super-ohmic bath. We assume
a small relaxation rate $\gamma \ll \Gamma$ for the bosonic modes.

In the adiabatic limit $\Omega \ll {\rm min}[\Gamma(t),\gamma]$ the
pumping current $I_\alpha(t)$ from the lead $\alpha$ to the quantum
dot can be cast as the sum of a first term, which we attribute to
elastic processes and a second, purely interacting term accounting
for inelasticity,
$I_{\alpha}(t)=I_{\alpha}^{el}(t)+I_{\alpha}^{in}(t)$ . The first
\be I^{el}_{\alpha}(t)=\sum_\beta \int \frac{d\w}{2\pi} (-f') \;{\rm
Im} \left[ { S}_{\alpha \beta} \de_t { S}^\dagger_{\beta \alpha}
\right], \label{I_el} \ee is analogous to Brouwer's
formula~\cite{Brouwer}, though written in terms of the instantaneous
\it elastic \rm S-matrix of the QD. Indeed, while here $f(\w)$ is
the Fermi function of the leads, ${ S}(\w,t)$ is defined as the
Wigner transform ${ S}(\w,t)=\int d\tau e^{i\w \tau} { S}(t+\tau /2,
t-\tau/2)$ of the time dependent scattering matrix \be {
S}_{\alpha,\beta}(t,t')=\delta_{\alpha \beta}\delta(t-t') -i 2\pi\nu
V_\alpha^\ast(t)\,G^r(t,t')\,V_\beta(t'), \ee where $G^r$ is the
full QD Green's function~\cite{Haugh}. When $\Omega \ll
\Gamma,\gamma$, we may express using the Fisher-Lee relation ${
S}_{\alpha,\beta}(\omega,t)=\delta_{\alpha \beta}-i2\,\pi\,\nu\,
V_\alpha^\ast(t)\,G^r(\omega,t)\, V_\beta(t)$ in terms of the
instantaneous QD Green's function $G^r(\omega,t)$, solving the time
independent problem with frozen parameters
$\{\epsilon(t),V(t)\}$~\cite{Arrachea}. Following
Ref.[\onlinecite{Langreth}], the latter is easily seen to be the
expression of the instantaneous elastic S-matrix .

The second term $I_{\alpha}^{in}$, in its most general form given in
Eq.(\ref{general1})-(\ref{general2}), is associated to the effect of
\it inelastic \rm scattering on the pumped current. This statement
is corroborated by the solution of the present problem at low
temperatures ($T\ll\Gamma$), where second order perturbation theory
in the QD-bath coupling applies. Indeed, assuming for simplicity
real QD-lead couplings we obtain \be
I^{in}_\alpha(t)=4\sum_\beta\int\frac{d\w}{2\pi} (-f')
\frac{\Gamma^{in}(\w,t)}{\Gamma(t)}
 {\rm Im}\left[ { T_0}^r_{\alpha \beta} \de_t {
T_0}^a_{\beta \alpha} \right], \label{I_in} \ee where the
noninteracting T-matrices ${ T_0}^{r,a}$ are defined by the
Fisher-Lee relation as
\begin{eqnarray}
{ T_0}^{r,a}_{\alpha,\beta}(\omega,t)=2\,\pi\,\nu\,
V_\alpha^\ast(t)\,G^{r,a}_0(\omega,t)\,V_\beta(t),
\end{eqnarray}
in terms of the instantaneous advanced/retarded QD Green's function
$G_{0}^{a,r}(\w,t)=(\w-\e(t)\mp \frac i 2 \Gamma(t))^{-1}$. Notice
that $I^{in}$ contains explicitly the inelastic scattering rate
associated to the coupling to the environment \be
\Gamma^{in}\!=\!\int\! \frac{d\e}{2} [J(\e)\!-\!J(-\e)]A(\w+\e,t)\{
f(\w+\e)+N(\e)\}, \ee where $A=-2\;{ \rm Im }[G^r_0]$, and $N(\e)$
is the Bose distribution function. As a consequence, this term plays
no role at $T=0$ as a result of the identity $\Gamma^{in} (\w=0)=0$.
The limit of weak inelasticity/dephasing considered here amounts to
$\Gamma^{in} \ll \Gamma$.

\begin{figure}
%\centering
%\psfrag{A}[cc][l]{$\delta \Gamma$}
%\psfrag{B}[lc][l]{$\hg$}
%\psfrag{C}[rc][l]{$-\hg$}
%\psfrag{F}[cc][l]{$\e_0$}
%\psfrag{E}[cc][l]{$\e$}
%\psfrag{G}[cc][l]{$-\e_0$}
\includegraphics[width=0.30\textwidth]{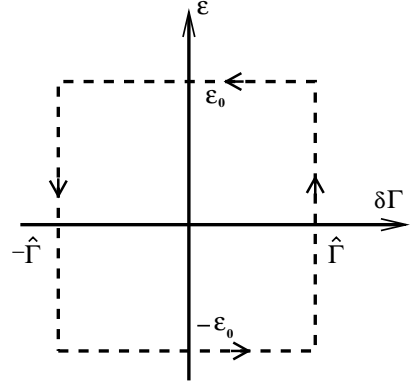}
\caption {The turnstile pumping cycle.} \label{fig_cycle}
\end{figure}

In order to illustrate the role of inelastic processes in a concrete
pumping cycle, let us consider the turnstile cycle of
Fig.\ref{fig_cycle}: keeping the width of the level constant in time
$\Gamma (t)=2\hg$, the pumping parameters are chosen to be the
energy level $\e \in [-\e_0, \e_0]$ and $\delta \Gamma
=\frac{\Gamma_L - \Gamma_R }{2}\in [-\hg, \hg]$. This cycle consists
in a periodic opening/closing of the coupling to the left/right
leads, followed by an inversion of the position of the level
$\epsilon$ when the dot is coupled to one lead. In the absence of
interaction and at low temperatures $T\ll \hat{\Gamma}$ the pumped
charge is \bea Q_0= \frac{2}{\pi} \left[ {\rm Arctg}(x)+
\frac{x}{x^2+1}\right] - \frac{8 \pi\,T^2}{3\hg^2}
\frac{x}{\left[x^2+1\right]^3},\label{noninteracting}
%T^2}{3}
%\frac{{\e_0} \hg^3}
%{\left[{\e_0}^2+\hg^2\right]^3},
\eea where $x=\e_0/\hg$ and the temperature dependence is due to
thermal broadening only. Notice that, for a noninteracting resonant
level, $Q_0 \simeq 1$ for $x \gg 1$.

Let us now account for interactions and clarify their effect on both
the elastic and the inelastic channels by separating their
contributions to the pumped charge.  The pumped charge associated to
Eq.(\ref{I_el}) is $Q^{el}=Q_0+\delta Q^{el}$ consists of the sum of
the noninteracting result Eq.(\ref{noninteracting}), plus a
correction originating from the loss of unitarity of the elastic
S-matrices. This second contribution, obtained by expanding
Eq.(\ref{I_el}) to second order in the QD-bath coupling , reads \bea
\delta Q^{el}\!=\!\frac {-\pi \alpha_s C T^{s+1}}{6\,\hg^2} \!
%\pi 6\, \alpha_s \frac{C T^{s+1}}{\hg^2}
\left[\frac{4x(9+8x^2+3x^4)}{(1+x^2)^2}\!+\! 12 {\rm
Arctg}(x)\right]\!, \label{brouw_1} \eea where we set for simplicity
$\gamma\!\ll\!T\!\ll\!\hg$, and $\alpha_s~=~\int  {d y}(3/
{8\,\pi^2}) \,\, \sech (\frac y 2)^2 \, \{-\Gamma(s+1)
\big(\li_{s+1} (-e^y)+\li_{s+1}(-e^{-y})\big)+ (3/ {\pi^2})
\Gamma(1+s)\, \li _{1+s}(1) \}$, $\li_s (z)$ being a polylogarithm
and $\Gamma(z)$ the Gamma function. Notice that summing up both
contributions we obtain \be Q^{el} \simeq 1-\pi^2 \alpha_s \frac{C
T^{s+1}}{\hg^2} \;\;\;\;\;\;\;\;{\rm for}\;\;x \gg 1 .\ee

Intuitively, the growth of the inelastic component should be at the
expense of the elastic one. In other words, $\delta Q^{el}$ and
$Q^{in}$ should have opposite sign. The presence of inelastic
channels, not accounted for in the elastic S-matrix, is also behind
the lack of quantization of $Q^{el}$ for $x \gg 1$. Nonetheless, a
proper account of inelastic scattering processes \it restores
quantization \rm for the \it total \rm pumped charge
$Q=Q^{el}+Q^{in}$, where $Q^{in}=\oint dt I^{in}$. Indeed, an
computation gives $Q^{in}=\frac \pi 6\, \alpha_s \frac{C
T^{s+1}}{\hg^2} \left[\frac{4x(5+3x^2)}{(1+x^2)^2} + 12 {\rm
Arctg}(x) \right]$. Therefore, summing up all contributions we have
\bea && Q= Q_0
%\frac 2 \pi \left(\textrm{Arctg}(x)+ \frac{{1}}{x^2+1}\right)
%+
%\nonumber
%\\ &&
 - \frac{8 \pi }{3}
\frac{x} {\left[x^2+1\right]^3} \frac{\left(\alpha_s\,C\,T^{s+1}
\right)}{\hat{\Gamma}^2} , \label{Q_final} \eea which is
approximately one for $x \gg 1$. This statement is valid also in
other temperature ranges (e.g. $T\ll \gamma, \hat{\Gamma}$).
Physically, as long as the resonance in the dot can still be fully
loaded and unloaded ($x \gg 1$) charge quantization
persists~\cite{Slava}, while for intermediate $x$ the inelastic
broadening of the resonance has a net effect on the temperature
dependence of $Q$.

\it General formalism \rm- The importance of inelastic channels and
of keeping track of unitarity pertain also to the general pumping
formula from which Eq.(\ref{I_in}) was obtained. Let us sketch its
derivation. Maintaining full generality, we consider a generic
multilevel, interacting quantum dot described by \bea
&&\ha=\sum_{k,\alpha} \epsilon_k\, c^\dagger_{k \alpha}\,c_{k
\alpha}+\sum_i \epsilon_i(t)\, d^\dagger_i\,
d_i+\nonumber \\
&&+\sum_{k,\alpha,i}\left[V_{\alpha,i} (t)\, d^\dagger_i\, c_{k
\alpha} +
  h.c.\right]+{\cal H}_{int}[d_i,d_i^\dagger].
\eea Here $\ha_{int}$ does not involve electrons in the leads.
Moreover, in order to be able to take the adiabatic limit, we assume
the instantaneous state of the interacting system to be
Fermi-liquid-like, characterized by an exponential decay of
instantaneous dot Green's functions with a typical rate
$\tilde{\Gamma}$.

The average current flowing through the dot may be  expressed in
terms of Keldysh Green's functions. Indeed, using \bea
\!\!I_\alpha(t)=\frac d {dt} \langle N_\alpha(t) \rangle= 2 \;{\rm
Re} \!\left[\!i\sum_{k, i} V_{\alpha,i}(t) \langle d^\dagger_i(t)
c_{k \alpha}(t) \rangle \right] \eea where $N_\alpha=\sum_k
c^\dagger_{k\alpha} c_{k \alpha}$, and using standard
techniques~\cite{Jauho} it is possible to show that \bea
&&I_\alpha (t)=i\big( f(t-t_1)\otimes { T}_{\alpha \alpha}^a(t_1,t)+\nonumber \\
&&- { T}_{\alpha \alpha}^r(t,t_1)\otimes f(t_1-t)\big)-i\,{
T}_{\alpha \alpha}^< (t,t) \label{I_st}. \eea Here $f(t)$ is the
Fourier transform of the Fermi function of the leads and the symbol
$\otimes$ stands for a convolution $A(t,t_1)\otimes B(t_1,t')=\int
dt_1 A(t,t_1) B(t_1,t')$. The full T-matrices are defined as
\begin{eqnarray}
{ T}^{r,a,<}_{\alpha,\beta}(t,t')=2\pi\nu {\bf
V}^*_{\alpha}(t)\;{\bf G}^{r,a,<}(t,t')\;{\bf V}_{\beta}(t')
\end{eqnarray}
where ${\bf G}^{r,a,<}_{i,j}$ are the full Green's functions of the
QD levels~\cite{Haugh}, and the boldface notation implies summation
over level indices. Information on the distribution function of the
dot, and of its deviation from the equilibrium $f(\omega)$ is
contained in ${\bf G}^<$.

A compact formula for the pumped current may be obtained expressing
${T}^<$ in terms of the QD self energy ${\bf \Sigma}^<={ \bf
\Sigma}_{0}^{<}+{\bf \Sigma}^<_{\rm int}$, where $[{\bf
\Sigma}_{0}^{<}(t,t')]_{i,j}=i \sum_{\alpha,k} V_{\alpha,i} (t)
f(\e_k) V^\ast_{\alpha,j}(t')\exp[-i\e_k(t-t')]$ is the self energy
associated to the coupling to the leads only, and ${\bf
\Sigma}_{int}$ is the one containing interaction (as well as
tunneling) vertices. Using the relation ${\bf G}^<={\bf G}^r \otimes
{\bf \Sigma}^< \otimes {\bf G}^a$ we may recast the current
(\ref{I_st}) in the form
%\begin{widetext}
\bea I_\alpha (t)&=&i\big( f\otimes { T}_{\alpha \alpha}^a- {
T}_{\alpha \alpha}^r\otimes f\big)+\sum_\gamma {
T}^r_{\alpha\,\gamma}\otimes f \otimes { T}^a_{\gamma \, \beta}
\nonumber \\ &-& i\,2\, \pi \, \nu \left({\bf V}_\alpha^\ast\,{\bf
G}^r\right) \otimes {\bf \Sigma}^<_{int} \otimes \left({{\bf G}^a
\bf V}_{\alpha} \right), \label{I_int} \eea
%\end{widetext}
where the time dependence of the r.h.s is analogous in structure to
Eq.(\ref{I_st}).

We are now in the position to take the adiabatic limit by performing
a gradient expansion~\cite{Vavilov,Haugh}. To this end, we notice
that we may write $I_{\alpha}(t)=\int d\w/(2\pi)\;I(\w,t)$, where
$I(\w,t)$ is the Wigner transform of the function on the r.h.s. of
Eq.(\ref{I_int}). In order to express it in terms of the
instantaneous quantities
%(i.e. $T^{r,a,<}(\w,t)$, $f(\w)$, etc..)
we recall that Wigner transform of a convolution has the simple
expansion~\cite{Haugh} \bea [A\otimes B]_{\w,\tau}&=& AB+\frac{1}{2
i}\left
  (\dt A\;\dw B-\dw A\dt B \right)+ \dots
\label{gexp} \eea In the adiabatic limit, the expansion may be
truncated to lowest nonvanishing order. Indeed, each derivative
produces a factor proportional to $\Omega/{\tilde{\Gamma}}\ll 1$.

It is crucial to note that, in order to expand Eq.(\ref{I_int})
consistently one has to account for conservation of probability or
unitarity. In equilibrium this is guaranteed by the optical theorem,
which out of equilibrium generalizes to  \bea
&&\!\!\!\!\!\!\!\!\!\!\!\! T^a_{\alpha \, \beta}(t,\,t')-T^r_{\alpha
\, \beta}(t,\,t')=i\, \sum_{\gamma}
T^r_{\alpha \, \gamma}\otimes  T^a_{\gamma \, \beta}+  \nonumber \\
&&\!\!\!\!\!\!\!\!\!\!\!\! +2\,\pi\,\nu \left({\bf V}^\ast_{\alpha}
{\bf G}^r \right) \otimes \left[{\bf \Sigma}^a_{int}- {\bf
\Sigma}^r_{int} \right] \otimes \left({{\bf G}^a \bf V}_{\alpha}
\right),~\label{unitary} \eea as one can easily see using the Dyson
equations. In the absence of interaction Eq.(\ref{unitary}) reduces
to the unitarity condition for the time dependent S-matrix,
$\sum_{\gamma}S_{\alpha,\gamma}\otimes
S^{\dagger}_{\gamma,\beta}=\delta_{\alpha,\beta}\delta(t-t')$.

Expanding now Eq.(\ref{I_int}) to first order in the gradients one
obtains two terms: the first, proportional to the Fermi function
$f(\w)$, vanishes once we take into account consistently the optical
theorem Eq.(\ref{unitary}) expanded to the same order in the
gradients. The remaining expression, which physically originates
from the deviation of the dot distribution function from the
equilibrium form, takes the form $I=I^{el}+I^{in}$ were the first
term $I^{el}$ is given by Eq.(\ref{I_el}), and
$I^{in}=I^{in}_1+I^{in}_2$, where
%\begin{widetext}
\bea \label{general1} I^{in}_{1}&=&i\nu\!\! \int\!\! d\w \;f'(\w)\;
\left[ {\bf \Sigma}_{int}^r(\w,t)\!-\!{\bf
\Sigma}_{int}^a(\w,t)\right]\;
 \\
 &&{\rm Im}\!\left[ \partial_t\left({\bf V}^\ast_\alpha(t){\bf
G^r}(\w,t)\right)
{\bf G^a}(\w,t){\bf V}_\alpha(t) \right] \nonumber \\
I^{in}_{2} &=&i\,\! \nu\!\int \!d \w \!\,{\bf
V}^\ast_{\alpha}(t)\,{\bf G^r}(\w,t)\!\Bigg[\bigg[ ({\bf
\Sigma}^a_{int})^{\prime}(\w,t)\!  \\ &-&\!({\bf
\Sigma}^r_{int})^{\prime}(\w,t) \bigg]\! f(\w)\!-\! ({\bf
\Sigma}^<_{int})^{\prime}(\w,t)\!\Bigg]\!{\bf G^a}(\w,t)\ {\bf
V}_{\alpha}(t)\label{general2} \nonumber
   \eea
%\end{widetext}

The equations above, together with Eq.(\ref{I_el}), reduce the
computation of the pumped current to that of the instantaneous
interacting self energies/Green's functions, ${\bf
\Sigma}^{r,a}_{int}$ and ${\bf G}^{r,a}$, and of their first order
expansion in gradients $({\bf \Sigma}^{r,a,<}_{int})^{\prime}$.
% functions using the standard tools of many-body theory.
% Here the subscript
%$[1]$ stresses the fact that the self energies appearing in
%$I^{in}_2$ are \it not \rm the instantaneous ones: when re-expressed
%in terms of instantaneous Green's functions, they should contain
%also terms of first order in the time derivatives (see example
%below). The zeroth order term, indeed, vanishes due to the
%fluctuation-dissipation relation
%$\left(\Sigma^a-\Sigma^r\right)f=\Sigma^<$ satisfied by the
%instantaneous self-energies. All other quantities appearing in
%q.(\ref{general}), Green's functions and self energies, are to be
%understood as instantaneous ones.
% and Eq.(\ref{general}) are the most general result
%of our work: it reduces the computation of the pumped current to
%that of the instantaneous interacting self energies/Green's
%functions using the standard tools of many-body theory.
This quantities may be computed using the standard tools of many
body theory within a perturbative scheme.
%These quantities may be computed using the standard tools of many
%body theory.
 For example,  the quantity
$({\bf \Sigma}^{r,a,<}_{int})^{'}$ may be easily obtained by \it
i~\rm) deriving, e.g. diagrammatically, an expression of the full
time dependent  ${\bf \Sigma}^{r,a,<}_{int}(t,t')$ in terms of the
Green's functions, \it ii~\rm) expanding the latter in gradients
using Eq.(\ref{gexp}) and isolating the term containing a single
time derivative.

To illustrate this in more detail, let us consider the calculation $
\Sigma_{int}$ to second order in perturbation theory for the
Hamiltonian Eq.(\ref{H_ph}). Neglecting the contribution of tadpole
diagrams, which for $\Omega \ll \gamma$ can be incorporated from the
beginning in the definition of $\e(t)$, the self energy reduces to
$\Sigma_{int}^{\lessgtr}(t,t')=i \sum_q \lambda_q^2
G_0^{\lessgtr}(t,t') D_q^{\lessgtr}(t,t')$ , where $G^{\lessgtr}_0$
are the noninteracting Green's functions of the dot, and
$D^{\lessgtr}_{q}(t,t')=-2\pi i( N(\w_q)\exp(\mp i
\w_q(t-t'))+(1+N(\w_q))\exp(\pm i \w_q(t-t'))$ are the bosonic
propagators. In order to expand these self energies in gradients,
one needs to first express $G^{\lessgtr}_0(t,t')=G_0^r(t,t_1)\otimes
\Sigma^{\lessgtr}_0(t_1,t_2) \otimes G_0^a(t_2,t')$, where
$\Sigma_0^{\lessgtr}(t_1,t_2)=\pm i 2 \pi \nu \sum_\alpha
V_\alpha(t_1) f(\pm t_1 \mp t_2) V^\ast(t_2)$, and then take the
Wigner transform of $\Sigma_{int}^{\lessgtr}(t,t')$.  Using the
formula for the gradient expansion and substituting in
Eq.(\ref{general1})-(\ref{general2}) we find, for real $V$,
Eq.(\ref{I_in}).

\it Relation to previous works \rm - Finally, we notice that a
direct comparison shows that the formula obtained in
Ref.\cite{Fazio} corresponds to the sum of $I^{el}+I^{in}_1$.
Therefore, the additional term $I^{in}_2$ corresponds to the
``irreducible vertex corrections'' for the emissivity reported in
Ref.\cite{Sela,Sela2} or to the corrections to the average time
approximation in Ref.\cite{Fazio2}.
%Notice that, while $I^{in}_2=0$ for the peculiar problem considered
%in Ref.\cite{Fazio}, $I^{in}_2$ cannot be seen in general as a small
%correction to $I^{in}_{1}$, the two being parametrically similar as
%shown above~\cite{Sela2}. We notice also that our expressions for
%$I^{in}_{1,2}$ are of easy access for the solution of practical
%problems, since they avoid the introduction of functional
%derivatives or vertex corrections required in
%Ref.\cite{Fazio,Fazio2,Sela}.
%The only limitation of our
%formalism is the requirement to extract the self energies with
%respect to a noninteracting Hamiltonian: an alternative approach
%studying pumping in a quantum dot in perturbation theory in the
%tunneling, but accounting exactly for the interaction, has been
%recently formulated in Ref.\cite{Konig2}.
The formalism presented in this letter requires the extraction of
the self energies with respect to a quadratic Hamiltonian. This does
not constraint future applications to weakly interacting problems
only: many interesting problems with strong correlations (e.g. Kondo
effects) can be mapped into quadratic models plus deviations from
them (e.g. by slave bosons, bosonization, etc..). An alterative
approach studying pumping in a quantum dot in perturbation theory in
the tunneling, but accounting exactly for the charging interaction,
has been recently formulated in Ref.\cite{Konig2}.

\it Conclusions \rm - In this paper, we derived a useful formula for
the pumped current through interacting quantum dots within a
perturbative scheme and used it to demonstrate that inelastic
scattering is the main effect forcing a description of pumping
beyond Brouwer's formula. Interestingly, even in the presence of
inelasticity, pumping can be still described
geometrically~\cite{Sela}. This statement draws a clear analogy
between pumping and the dissipative Berry Phase
problem~\cite{Gefen}, whose implications, in view of recent work
connecting Brouwer's formula to geometric phases~\cite{Ho}, are
worth of further investigation.

We thank R. Fazio, R. Ferrari, L. Molinari, Y. Oreg, E. Sela, A.
Schiller, J. von Delft for enlightening discussions. D.F. would like
to thank the Stat. Phys. group at SISSA for hospitality and support
through a SISSA undergraduate fellowship.

\end{document}